\documentclass[]{spie}  

\newcommand*\aap{A\&A}

\newcommand*\ao{Appl Opt}

\newcommand*\apj{ApJ}
\newcommand*\apjl{ApJ}

\newcommand*\apss{Ap\&SS}
\newcommand*\araa{ARA\&A}

\newcommand*\icarus{Icarus}

\newcommand*\nat{Nature}

\newcommand*\pasp{PASP}

\usepackage{amsmath,amsfonts,amssymb}
\usepackage{graphicx}

\title{Characterizing the atmosphere of Proxima b with a space-based mid-infrared nulling interferometer}

\author[a]{D. Defr\`ere}
\author[b]{A. L\'eger}
\author[a]{O. Absil}
\author[c]{A. Garcia Munoz}
\author[d]{J.L. Grenfell}
\author[c]{M. Godolt}
\author[a]{J. Loicq}
\author[e]{J. Kammerer}
\author[f]{S. Quanz}
\author[b]{H. Rauer}
\author[a]{L. Schifano}
\author[g]{F. Tian}

\affil[a]{Space sciences, Technologies \& Astrophysics Research (STAR) Institute, Universit\'e de Li\`ege, 19c all\'ee du Six Aout, b\^at B5c, B-4000 Li\`ege, Belgium}
\affil[b]{Institut d'Astrophysique Spatiale, Universit\'e de Paris-Sud, Orsay, France}
\affil[c]{Technische Universit\"at Berlin, Berlin, Germany}
\affil[d]{Institute for Planetary Research, German Aerospace Center, Berlin, Germany}
\affil[e]{Research School of Astronomy \& Astrophysics, Australian National University, Canberra, ACT 2611, Australia}
\affil[f]{Eidgen\"ossische Technische Hochschule (ETH) Zurich, Institute for Particle Physics and Astrophysics, Zurich, Switzerland}
\affil[g]{Department of Earth System Science. Tsinghua University, Beijing, China}

\authorinfo{Denis Defr\`ere: ddefrere@uliege.be, +323669758}

\begin{document} 
\maketitle

\begin{abstract}
Proxima b is our nearest potentially rocky exoplanet and represents a formidable opportunity for exoplanet science and possibly astrobiology. With an angular separation of only 35~mas (or 0.05~AU) from its host star, Proxima b is however hardly observable with current imaging telescopes and future space-based coronagraphs. One way to separate the photons of the planet from those of its host star is to use an interferometer that can easily resolve such spatial scales. In addition, its proximity to Earth and its favorable contrast ratio compared with its host M dwarf (approximately 10$^{-5}$ at 10 microns) makes it an ideal target for a space-based nulling interferometer with relatively small apertures. In this paper, we present the motivation for observing this planet in the mid-infrared (5-20 microns) and the corresponding technological challenges. Then, we describe the concept of a space-based infrared interferometer with relatively small ($<$1m in diameter) apertures that can measure key details of Proxima b, such as its size, temperature, climate structure, as well as the presence of important atmospheric molecules such as H$_2$O, CO$_2$, O$_3$, and CH$_4$. Finally, we illustrate the concept by showing realistic observations using synthetic spectra of Proxima b computed with coupled climate chemistry models.  
\end{abstract}

\keywords{Exoplanet, Habitable, Proxima Cen, Proxima b, Nulling, Infrared imaging}

\section{INTRODUCTION}
\label{sec:intro}  

Since the discovery of the first exoplanet around a Sun-like star \cite{Mayor:1995}, the field of exoplanet research has been thriving and discoveries have been made at an accelerating pace. Today, more than 3700 exoplanets have been detected in our galaxy as well as approximately 4500 candidates, most of which are likely to be true planets. Although the current detection techniques show peculiar biases in terms of planetary sizes, orbital distances, or temperatures, it is already clear that the progressively emerging planet population presents an incredible diversity that goes well beyond what one can see in the Solar system. From young and hot to old and cold, from small and rocky to giant and gaseous, from tight to wide orbits and from circular to eccentric orbits, exoplanets cover a huge parameter space. While they represent only a small fraction of the known exoplanet population, rocky planets are very common, with 0.4 to 1.0 per star depending on e.g. stellar spectral type\cite{Winn:2015} and they cover a wide range of equilibrium temperatures. Nevertheless, very little is currently known about their atmospheres.

A promising case for making progress in this direction is Proxima b, a rocky planet that was recently discovered in the inner habitable zone of our nearest neighbouring star\cite{Anglada:2016}. Proxima b is a 1.3 Earth-mass planet orbiting at a distance of 0.05AU where it receives $\sim$65\% of the net insolation of the Earth. Studies using atmospheric column models\cite{Turbet:2016,Turbet:2018,Meadows:2018} suggest that the planet might be habitable if (1) it was formed further out (hence avoiding likely dessication during the early stages) or/and if (2) it possesses a thick, protective H$_2$ envelope from the protoplanetary disk during its formation. The weak insolation suggests that a strong greenhouse effect would be needed to maintain surface habitability e.g. via  a surface pressure of several bar of CO$_2$. 


Performing remote spectroscopy of this planet will provide a first insight on its atmosphere and provides a chance to assess its potential habitability. Since Proxima b is unlikely to transit, its atmospheric characterization will require direct imaging techniques or phase curve measurements\cite{Kreidberg:2016}. In this paper, we present and discuss an instrumental concept that would be able to take the mid-infrared (5-20 microns) spectrum of Proxima b. We present the key advantages of the mid-infrared regime in Section~\ref{sec:infrared_imaging} and why it is necessary to observe from space in Section~\ref{sec:space}. In Section~\ref{sec:cconcept}, we present the observing challenge, the measurement concept, and the instrumental requirements as well as a few illustrative simulated observations.

\section{Why the mid-infrared?} \label{sec:infrared_imaging}

\begin{figure}
\centering
\includegraphics[width=7.6cm]{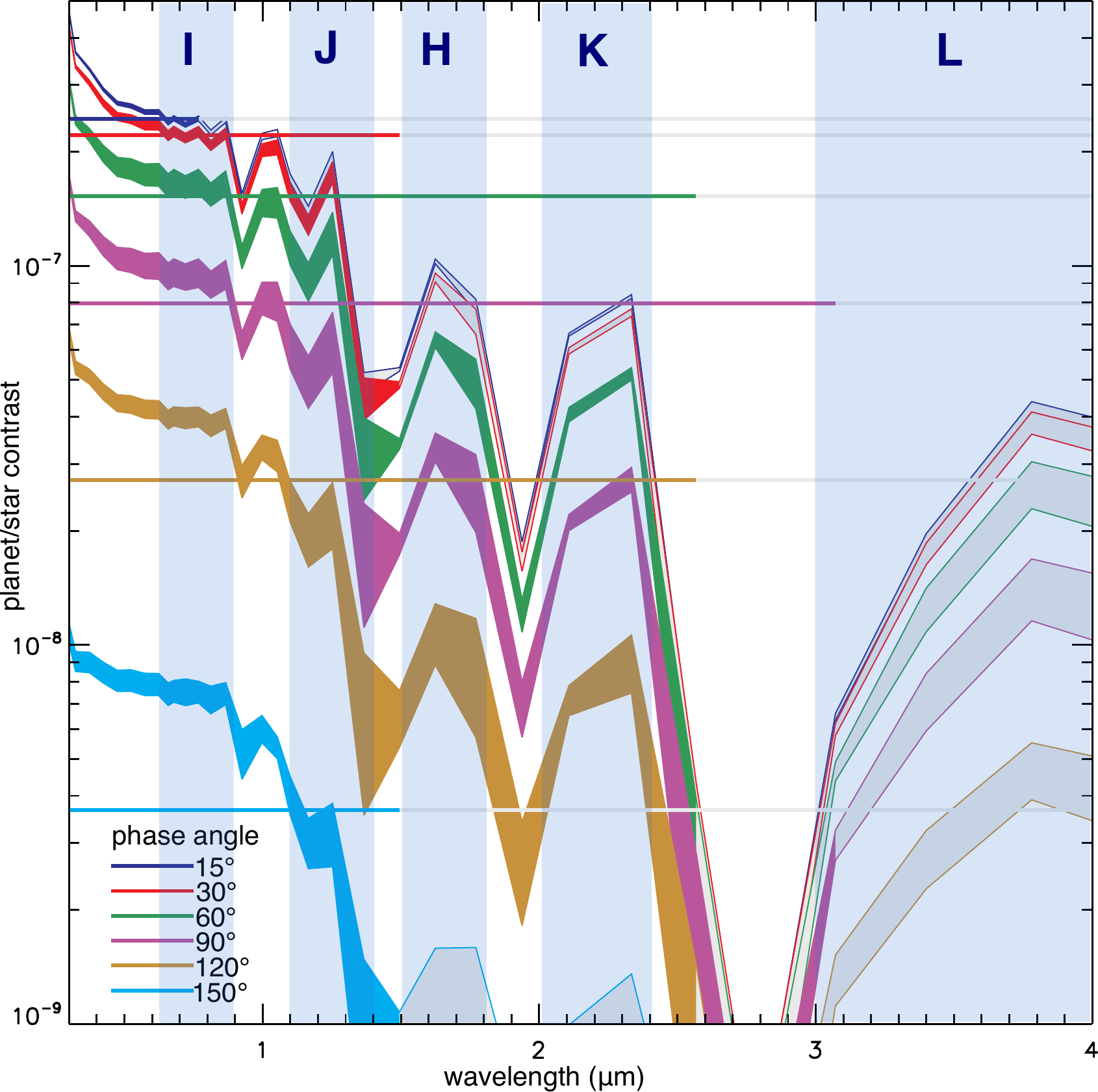}
\includegraphics[width=7.85cm]{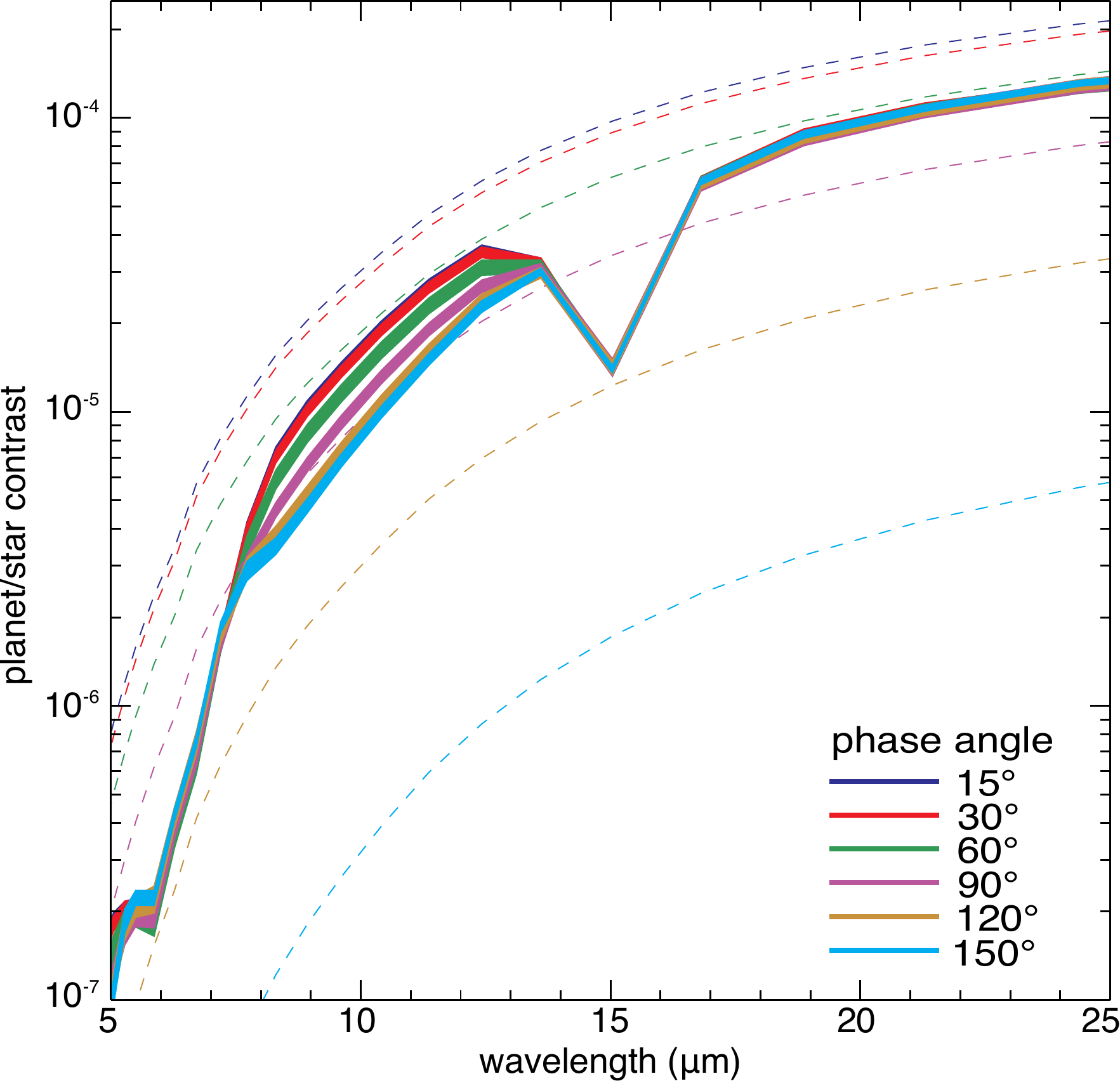}
\caption{Reflection (left) and emission (right) spectra of Proxima b assuming an Earth-like atmosphere (1 bar of N$_2$, 376 ppm of CO$_2$, a global ocean, but no O$_2$/O$_3$) showing that the planet/star contrast is more favorable in the mid-infrared\cite{Turbet:2016}. Each color corresponds to a different phase angle ($15^{\circ}$ meaning that the observer nearly looks at the substellar point and $90^{\circ}$ at a point on the terminator). The thickness of each curve indicates the range of possible values depending on the actual observing geometry. In the left plot, straight lines are calculated for a constant surface albedo of 0.4. Curves are plotted in grey when the angular separation falls below twice the diffraction limit of the ELT ($2\times1.2 \lambda/D$). In the right plot, dashed lines are calculated for a planet with no atmosphere with a constant surface albedo of 0.2. These plots are computed for a fixed planetary radius of 1.1~R$_\oplus$. Figure from Turbet al.\ 2016\cite{Turbet:2016}.}\label{fig:contrast_IR}
\end{figure}

Direct imaging in the mid-infrared has a key role to play in studying exoplanets and understanding their atmospheres. In addition to providing a favorable planet/star contrast to detect the emission of HZ exoplanets (see e.g.\ Figure~\ref{fig:contrast_IR} for Proxima b), this wavelength region also provides data to measure key planetary parameters, such as their size, temperature, presence of an atmosphere, climate structure, as well as the presence of important atmospheric molecules. These points are briefly discussed in the following:

\begin{itemize}
\item Presence of an atmosphere and basic planetary properties. Even without the use of spectroscopy, and for any atmospheric composition, monitoring the variations of thermal emission of an exoplanet during its orbital motion, in a few or even one single broad band, can fundamentally constrain the atmospheric mass and climate\cite{GomezLeal2016,ThesisIlleana}. 

\item Atmospheric composition. In addition to the constraints obtained from orbital broadband photometry, mid-infrared spectroscopy is crucial to refine the nature of the atmosphere by providing atmospheric species, constraining the temperature, pressure structure, cloudiness, and determining whether the planet could potentially harbor life. The mid-infrared regime contains several important spectral features (e.g., H$_2$O, CO$_2$, O$_3$, CH$_4$) that can be detected at low to medium spectral resolving power \cite{Desmarais:2002}.

\item Surface conditions and habitability. Surface temperature is a key property to study the habitability of an exoplanet and to search for life. For planets similar to Earth, a detailed study \cite{vonParis:2013} shows that surface conditions (temperature and pressure) can be characterized relatively well from mid-infrared observations (to within $\sim$10 K at 3-$\sigma$) with S/Ns between 10 and 30, depending on spectral resolution. For planets much warmer than Earth, the surface temperature might not be determined by infrared observations alone, since such warm conditions could precipitate a so-called runaway greenhouse that would anyway prevent the planet from maintaining habitability. 

\end{itemize}

Note that Proxima b will be directly observable with future telescopes but these observations will be limited to the visible regime from space (e.g. Habex/LUVOIR) and the near-infrared regime with ground-based 40-m class telescopes such as the ELT (assuming an inner working angle of 2$\times\lambda$/D, see review in Meadows et al.\ 2018\cite{Meadows:2018}). 

\section{Why in space?} \label{sec:space}

Observing stars at mid-infrared wavelengths through the Earth's atmosphere can be compared with observing candles from behind a wall of fire. To be specific, the atmosphere emits $2\times10^9$ ph/s/arcsec$^2$/$\mu$m/m$^2$ at 10~$\mu$m under standard conditions at Mauna Kea \cite{Lord:1992}. This is more than most stars and 10 billion times brighter than a 300K Earth-sized planet located at 10\,pc when observed with an 8-m telescope. Although advanced chopping/nodding techniques have been developed to remove the spatial and temporal fluctuations of the background, photon noise remains an unavoidable limitation. For instance, assuming an ELT-like aperture of 40~m and a bandwidth of 0.1~$\mu$m centered at 10~$\mu$m, it would require approximately 600 days to reach a S/N of 5 for a 300\,K planet located at 10\,pc based on pure photon noise considerations. 

Another major problem encountered by ground-based observatories arises due to  atmospheric turbulence. The water vapor component of the atmospheric seeing in particular has been (or is) a serious issue for high-precision instruments on the \emph{Keck Interferometer Nuller} \cite{Colavita:2009} or the \emph{Large Binocular Telescope Interferometer} \cite{Defrere:2016} and is expected to be a major issue for future instruments installed on ELTs like METIS \cite{Kendrew:2008}. Combined with background noise, these two effects have limited the performance achieved by ground-based instruments to approximately three orders of magnitudes below that required to tackle the scientific goal of this proposal.\\

Finally, because the Earth's atmosphere is mostly opaque at the wavelengths corresponding to major molecular absorptions (such as water vapour and carbon dioxide), searching for the broad spectral signatures of major molecular species in planetary atmospheres will generally be very difficult from the ground.

\section{Measurement concept}  \label{sec:cconcept}
\subsection{The observing challenge} \label{sec:challenge}

\begin{figure}[!t]
	\begin{center}
		\includegraphics[height=9.8 cm]{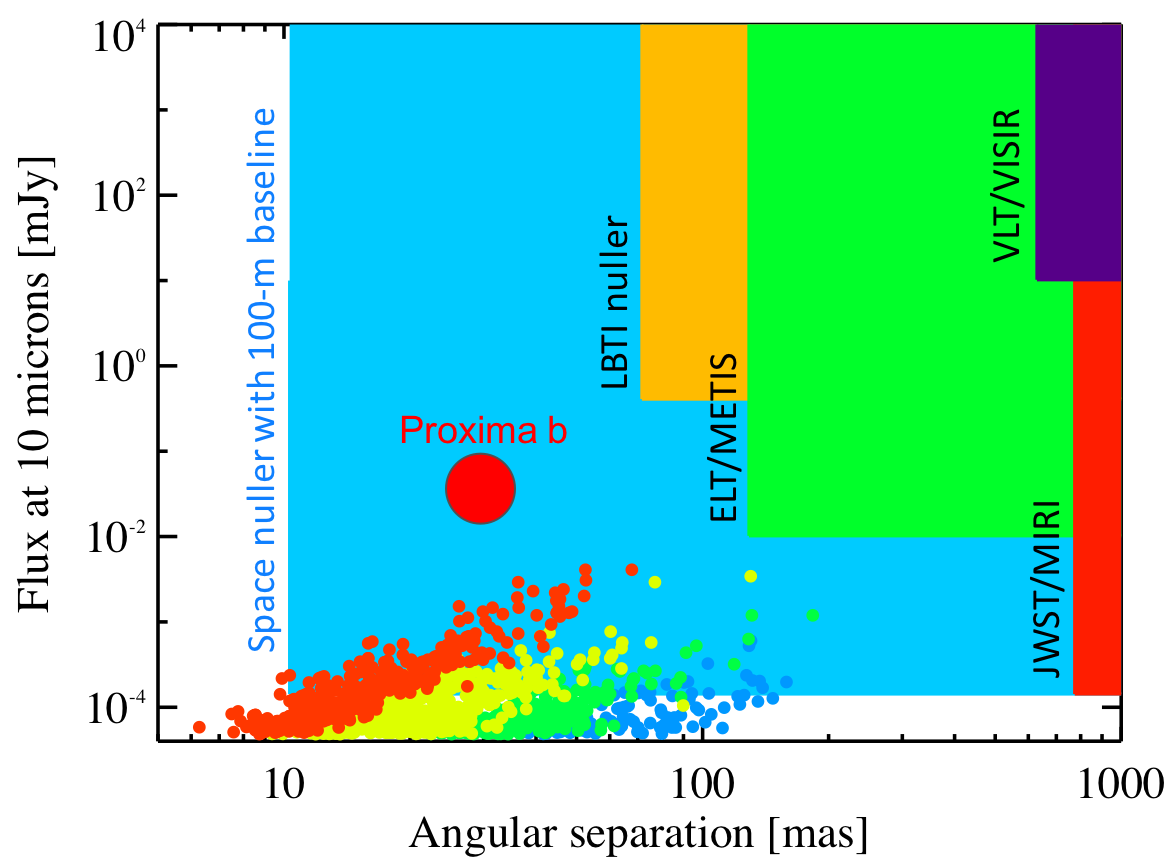}
		\caption{Flux at 10~$\mu$m as a function of angular separation of putative 300\,K blackbody planets located in the middle of the HZ of nearby single main-sequence stars (see colored circles: M stars in red, K stars in yellow, G stars in green, and F stars in blue, Darwin All Sky Star Catalog\cite{Kaltenegger:2010}). The 10-$\sigma$ sensitivity corresponding to one day of integration of current and planned mid-infrared instruments is indicated by the colored rectangles \cite{Brandl:2014} and compared to that of a four 2m-aperture space-based nulling instrument (see blue shaded area, assuming no exozodiacal dust). Note that this plot does not take into account aperture masking modes, which improve the angular resolution but reduce the sensitivity.}\label{fig:synergies}
		\vspace{-1.5em}
	\end{center}
\end{figure}

Obtaining high-quality infrared spectra is an essential requirement to study the atmosphere of exoplanets \cite{Burrows:2014}. However, directly detecting the photons from a rocky exoplanet is a very challenging observational task due to several factors, in particular the high contrast and the small angular separation between the planet and its host star. Figure~\ref{fig:synergies} shows that none of current or foreseen instruments have neither the necessary angular resolution nor the sensitivity to observe Proxima b at 10 $\mu$m, but also to survey the habitable zone of a sample of nearby main-sequence stars. In the case of the JWST, the impressive sensitivity provided by the large collecting area (25~m$^2$) and cold (40\,K) telescope optics can only be utilized by coronagraphs which are expected to achieve a best case contrast at 10.6~$\mu$m of 10$^{-4}$ to 10$^{-5}$, for separations larger than 0.5 to 1.0 arcsecond \cite{Boccaletti:2015}. With such performances, the detection of warm and young exo-Jupiters is likely the closest that the JWST/MIRI instrument can approach. For the ELT, the massive gain in collecting area (980~m$^2$) compared to the JWST offsets the impact of having warm optics to give comparable sensitivity limits for the METIS instrument, operating at 3 to 19~$\mu$m. METIS \cite{Brandl:2016} will be equipped with coronagraphs which can in principle achieve contrasts of $\sim$10$^{-7}$ at separations of $\sim$0.7 arcsecond necessary to directly image a putative exo-Earth orbiting $\alpha$~Cen as well as $\sim$10 small planets (1 to 4 R$_\oplus$) with equilibrium temperatures between 200 and 500\,K around the nearest stars \cite{Quanz15}. However, achieving this performance in practice at a ground-based observatory where image quality and stability are dependent on an advanced adaptive optics system, will be challenging. Furthermore, due to the scarcity of available photons, the measurement would be restricted to a photometric detection, with little hope of spectroscopic follow-up.


\subsection{Extracting the planetary photons}

\begin{figure}[!t]
	\begin{center}
\includegraphics[height=6.2cm]{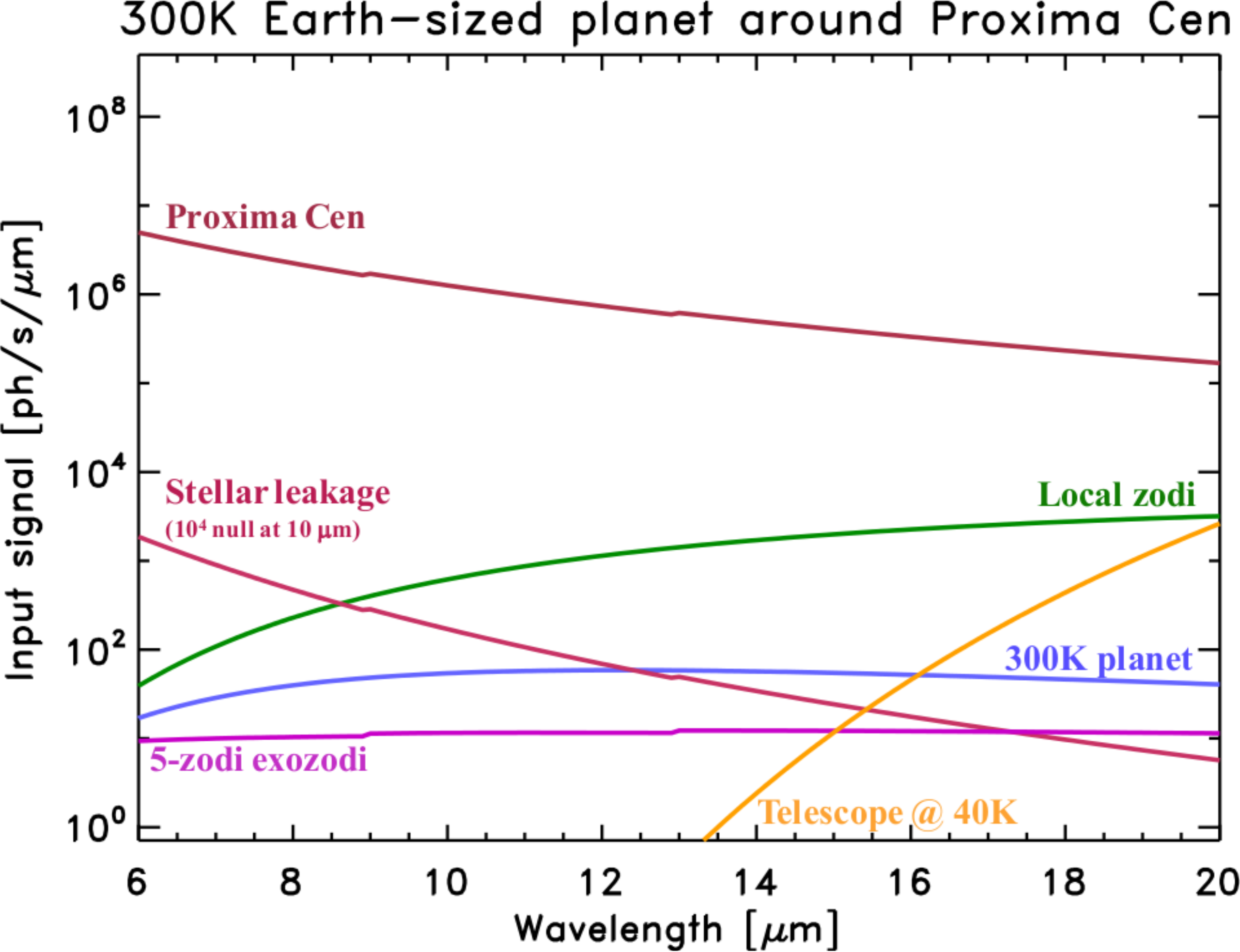}
        \includegraphics[height=6.2cm]{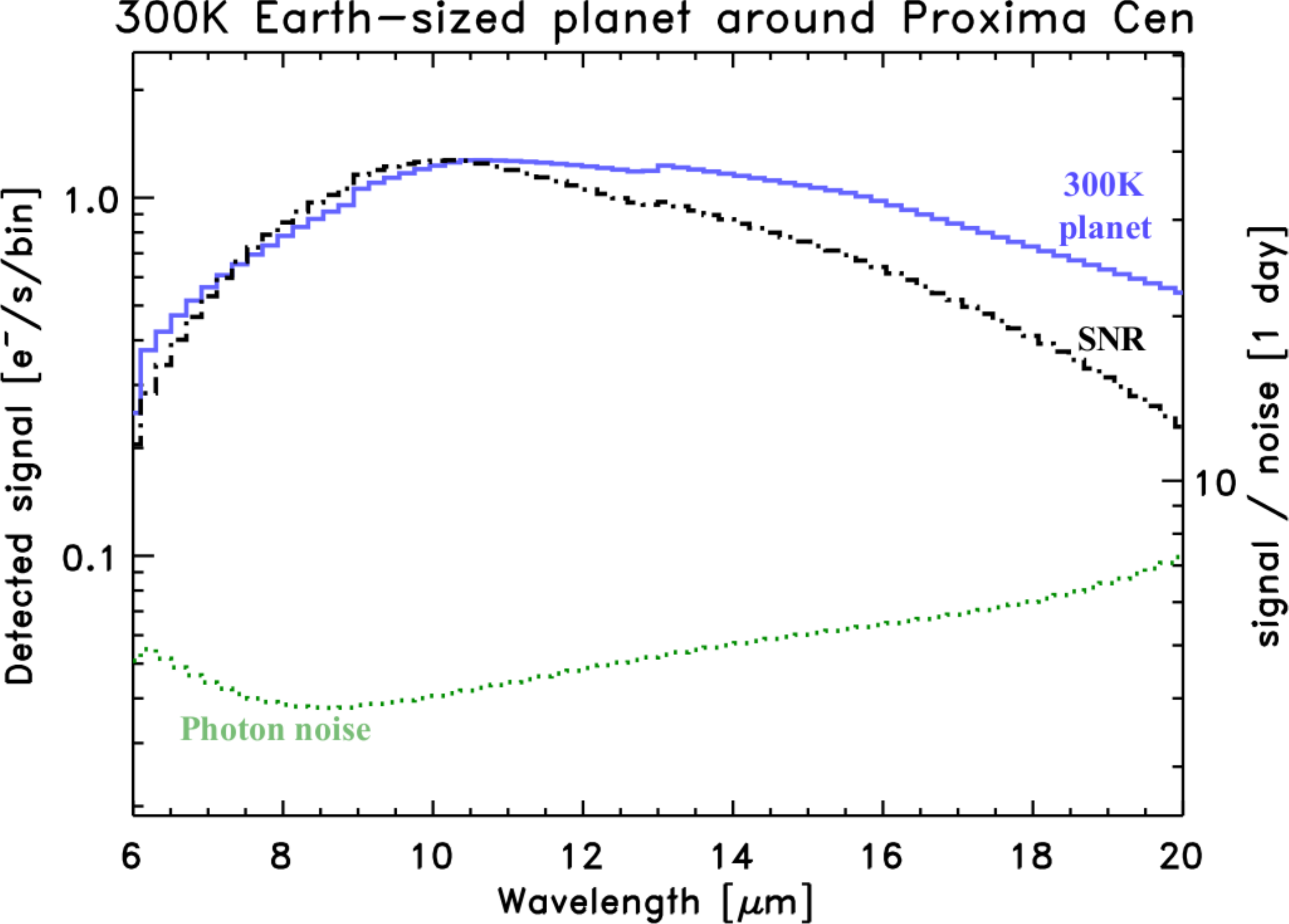}
		\caption{Left, different flux sources seen by a space-based infrared interferometer when observing a 300~K Earth-sized planet around Proxima Cen. Right, corresponding SNR obtained after one day of integration assuming a configuration with four 75-cm collecting mirrors and a mean spectral resolution of 40 (and taking into account only photon noise). More details on the flux sources, instrumental parameters and simulation software can be found in Defr\`ere et al.\ 2010\cite{Defrere:2010}.} \label{fig:challenge2}
		\vspace{-1.5em}
	\end{center}
\end{figure}

\begin{figure}[!t]
	\begin{center}
        \includegraphics[height=9.4cm]{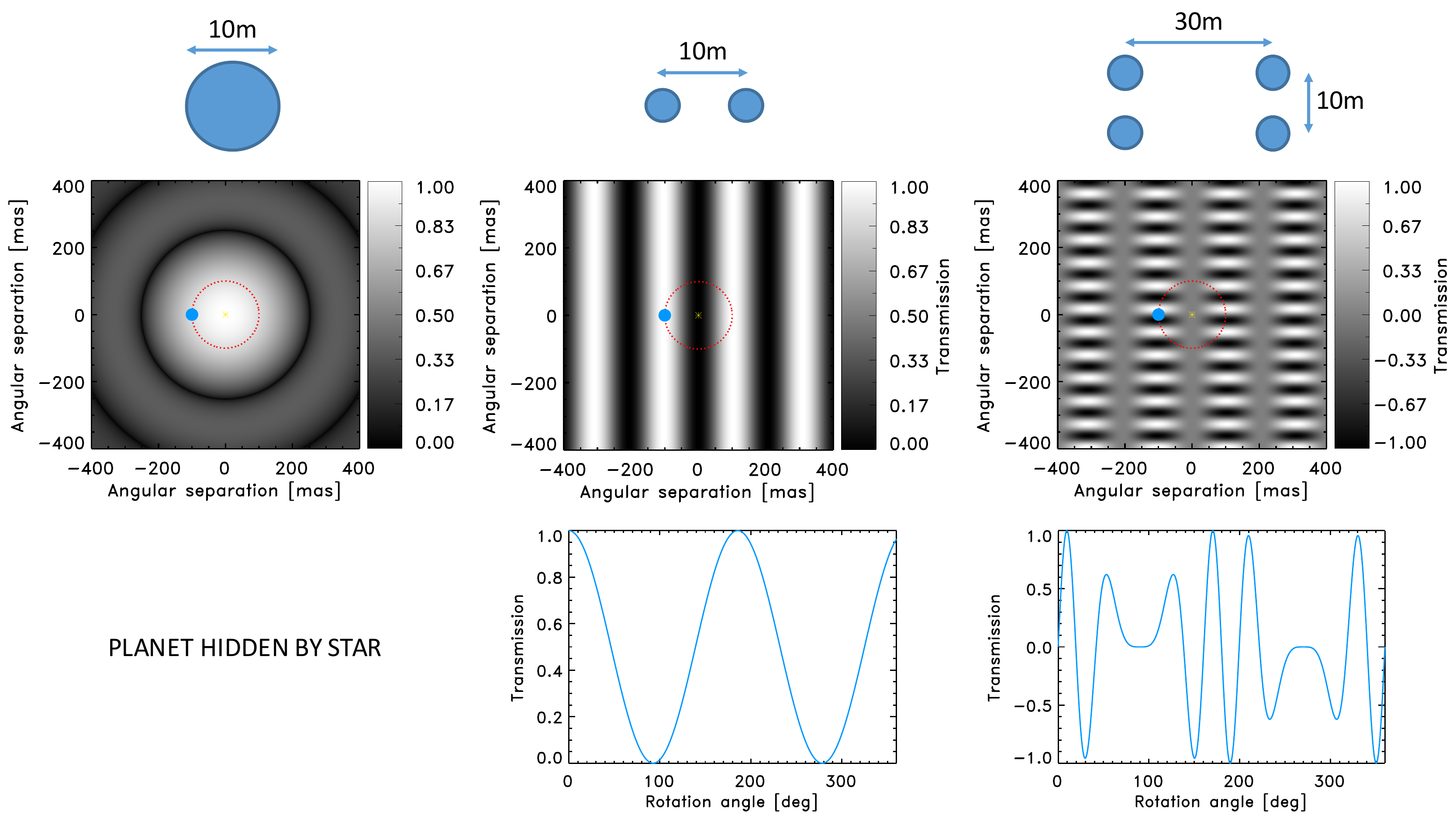}
		\caption{Planet orbiting at 1\,AU around a star at 10\,pc as seen by a single-aperture 10-m telescope (left), 10-m Bracewell interferometer (middle), and a 10x30m four-aperture nulling interferometer (right, all plots computed at 10\,$\mu$m). For the single-aperture 10-m telescope, the PSF is represented in the middle panel of the left row and shows that the planet is both unresolved and hidden by the much brighter star. For the interferometers, the planet is resolved and the stellar flux strongly reduced as shown by the transmission maps in the middle plots. This transmission map is effectively projected onto the sky, blocking some regions while transmitting others. The bottom panels show the modulation of the planet signal as the array rotates (it can be negative when the signals from two different sub-arrays are subtracted). The signal due to the star is nulled in first approximation, and appears as noise that is independent of the rotation angle. For a given rotation period, higher modulation frequencies can be obtained with the four-aperture interferometer, which provides a better performance against low-frequency instrumental drifts. This signal can then be deconvolved to obtain the planet signal \cite{Bracewell:1978,Leger:1996,Angel:1997} or to form an image of the planetary system \cite{Lay:2005}. Unless giant primary mirrors can be launched into space, nulling interferometry is mandatory to characterize in the mid-infrared the atmosphere and investigate the habitability of exoplanets.}\label{fig:modulation}
		\vspace{-1.5em}
	\end{center}
\end{figure}

A technique that can realistically tackle the observing challenge presented in the previous section is nulling interferometry, as initially proposed in 1978 to detect ``non-solar" planets \cite{Bracewell:1978}. The basic principle of this technique is to combine the beams coming from two telescopes in phase opposition so that a dark fringe appears on the line of sight (see middle panel of Figure~\ref{fig:modulation}), which strongly reduces the stellar emission. Considering the two-telescope interferometer initially proposed by Bracewell, the response on the plane of the sky is a series of sinusoidal fringes, with angular spacing of $\lambda/b$. By adjusting the baseline length ($b$) and orientation, the transmission of the off-axis planetary companion can then be maximized. However, even when the stellar emission is sufficiently reduced, it is generally not possible to detect Earth-like planets with a static array configuration, because their emission is dominated by the thermal contribution of warm dust in our solar system as well as around the target stars (see left panel of Figure~\ref{fig:challenge2}). This is the reason why Bracewell proposed to rotate the interferometer so that the planetary signal is modulated by alternatively crossing high and low transmission regions, while the stellar signal and the background emission remain constant (see bottom panels in Figure~\ref{fig:modulation}). The planetary signal can then be retrieved by synchronous demodulation. This modulation technique is in many ways similar to the use of a chopper wheel that allows the detection of infrared sources against a thermal background and/or drifting detector offsets. In addition, it was quickly realized that the array cannot be rotated sufficiently fast to mitigate low frequency instrumental drifts \cite{Lay:2004} and a number of interferometer configurations with more than two collectors have then been proposed to perform faster modulation and overcome this problem by using phase chopping \cite{Angel:1997,Mennesson:1997}. The principle of phase chopping is to synthesize two different transmission maps with the same telescope array, by applying different phase shifts in the beam combination process. By differencing two different transmission maps, it is possible to isolate the planetary signal from the contributions of the star, local zodiacal cloud, exozodiacal cloud, stray light, thermal, or detector gain. An example of the transmission map for a four-telescope interferometer is shown in the right panel of Figure~\ref{fig:modulation}. 



\subsection{Instrumental requirements}\label{sec:requirements}

The instrumental requirements for a direct imaging mid-infrared mission able to spectroscopically characterize the atmosphere of rocky exoplanets and, in particular, Proxima b have been recently reviewed by Defr\`ere et al. 2017\cite{Defrere:2017}. In short, given the baseline length required to observe Proxima b ($\sim$40~m at 10~$\mu$m), the top-level instrumental requirement is the ability to fly an array of telescopes in a coordinated way. Remarkable advances in technology have been made in Europe in recent years with the space-based demonstration of this technology by the PRISMA mission (http://www.snsb.se/en/Home/Space-Activities-in-Sweden/Satellites/). PRISMA demonstrated a sub-cm positioning accuracy between two spacecraft, mainly limited by the metrology system (GPS and RF). The launch of ESA's PROBA-3 mission at the end of 2020 will provide further valuable free-flyer positioning accuracy results (sub-mm), which exceeds the requirements for a space-based nulling interferometer. Extending the flight-tested building-block functionality from a distributed two-spacecraft instrument to an instrument with more than two spacecrafts mainly relies on the replication of the coordination functionality and does not present additional complexity in terms of procedures according to the PRISMA navigation team. While formation flying can then be considered to have reached TRL 9, once PROBA-3 has flown, an uncertainty remains regarding fuel usage and the possible lifetime of such a mission. 

Regarding starlight suppression, a considerable expertise has been developed on this topic over the past 20 years, both in academic and industrial centers across the globe. Approximately 35 PhD theses have been dedicated to this topic and more than 40 refereed papers. These efforts culminated with laboratory demonstrations at room temperature mainly at the \emph{Jet Propulsion Laboratory} (JPL) in the US. For instance, work  with  the  Adaptive  Nuller  has  indicated  that  mid-infrared nulls of 10$^{-5}$ are achievable with a bandwidth of 34\% and a mean wavelength of 10\,$\mu$m \cite{Peters:2010}. In addition, the planet detection testbed was developed in parallel and demonstrated the  main components of a high performance four-beam nulling interferometer at a level matching that needed for the space mission \cite{Martin:2010}. At 10~$\mu$m with 10\% bandwidth, this has achieved nulling of $8\times 10^{-6}$ (the flight requirement is $10^{-5}$), starlight suppression of $10^{-8}$ after post-processing, and actual planet detection at a planet-to-star contrast of $3\times 10^{-7}$, i.e., the Earth-Sun contrast. The phase chopping technique \cite{Mennesson:2005} has also been implemented and validated on-sky with the Keck Nuller Interferometer \cite{Colavita:2009}. A null stability of a few $\sim$10$^{-3}$ was achieved, mainly limited by the large thermal background and variable water vapor content, both effects specific to ground-based mid-infrared observations.


\subsection{Illustrative observations}

\begin{figure}[!t]
	\begin{center}
		\includegraphics[width=17.0cm]{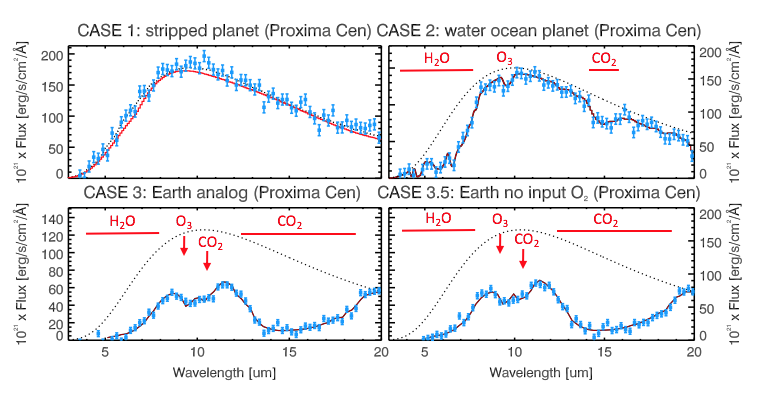}
		\caption{Simulated mid-infrared spectra of Proxima b with various atmospheric properties that can be studied by remote sensing (blue line, L\'eger et al. in prep) and corresponding blackbody emission for the planetary surface temperature (T$_{\rm s}$, dotted line). The synthetic spectra are computed by coupled climate chemistry models (Cases 1, 2, Rauer et al.\ 2011; Cases 3, 3.5, Tian et al.\ 2014). \underline{Case 1}: a rocky planet (M=4.0\,M$_\oplus$, R=1.5\,R$_\oplus$) in the HZ of Proxima Cen. The stellar insolation (S) is 65\% of that of Earth (S$_{\rm e}$) and T$_{\rm s}$=240\,K. No atmosphere (stripped planet), using Apollo Moon sample 15071 for IR emissivity. \underline{Case 2}: Water-ocean planet (M=2.0\,M$_\oplus$, R=1.5\,R$_\oplus$) in HZ of Proxima Cen (S=1.05S$_{\rm e}$) with arbitrary (possibly abiotic) O$_2$ input that could be due e.g. to strong escape. Atmosphere:  P$_{\rm N2}$=1\,bar, P$_{\rm O2}$=200\,mbar, 1\,ppm P$_{\rm CO2}$, saturated H$_2$O vapour, T$_{\rm s}$ = 290\,K, calculated O$_3$ with coupled chemistry. \underline{Case 3}: a large Earth-analog planet (M = 4.0 M$_\oplus$, R = 1.5 R$_\oplus$) in HZ of Proxima Cen (S=0.65S$_{\rm e}$), but with a strong CO$_2$ Greenhouse effect bringing T$_{\rm s}$ to 280\,K. Atmosphere: P$_{\rm CO2}$=300\,mbar, P$_{\rm N2}$=500\,mbar, P$_{\rm O2}$=200\,mbar (possibly biotic), H$_2$O from vapour pressure, calculated O$_3$ with coupled chemistry. \underline{Case 3.5}: a rocky planet in HZ of Proxima Cen (S=0.65S$_{\rm e}$), with P$_{\rm CO2}$=300\,mbar bringing T$_{\rm s}$ to 280\,K, H$_2$O from vapour pressure, no O$_2$ input, calculated O$_2$ and O$_3$ from coupled chemistry induced by the UV flux of the M star. Simulated observations (R=40) imposing a S/N of 20 on continuum detection at 10~$\mu$m are over-plotted as blue points. Besides O$_3$ in the atmosphere of the water ocean planet around Proxima Cen (Case 2), all spectral features can be retrieved in a single visit with these requirements (R=40 and S/N=20). Detecting O$_3$ in Case 2 would require a higher S/N or follow-up observations.} \label{fig:case_out}
		\vspace{-1.5em}
	\end{center}
\end{figure}

To illustrate typical observations obtained with a nulling interferometer, we show in Figure~\ref{fig:case_out} four examples of possible mid-infrared spectrum for Proxima b. These spectra have been computed using coupled climate chemistry models \cite{Rauer:2011,Tian:2014} and are briefly described here (more information can be found in L\'eger et al.\ in prep): a stripped planet without an atmosphere, a water ocean planet\cite{Leger:2004}, and two cases of large Earth-analog planets (one with possibly biotic O$_2$ and one without any O$_2$ input, see legend for more information). Each kind of planet presents a specific spectral signature that can be observed and studied remotely to inform us about the potential nature of Proxima b: the stripped planet (Case 1) does not show observable spectral features, the water ocean planet shows prominent H$_2$O features (Cases 2), and the Earth-analog planets present the triple signature\cite{Selsis:2002} seen in Earth's spectrum (H$_2$O, CO$_2$, and O$_3$, Cases 3 and 3.5). 

The accuracy with which these atmospheric parameters can be retrieved from mid-infrared spectra critically depends on the spectral resolution and S/N of the measured spectra. For instance, a detailed study by von Paris et al.\cite{vonParis:2013} concludes that a mid-infrared spectrum with a resolution of 25 is sufficient to retrieve the surface conditions (temperature and pressure) of an Earth-like planet and to infer the presence of CO$_2$ but show that the detection of O$_3$ and H$_2$O would be marginal, even at high S/N. A promising way to improve the parameter retrieval is to monitor the variation of the spectra with phase angle (which would require repeated observations). This approach has never been tested because it requires detailed 3D models with coupled photochemistry (for O$_3$) but would provide strong additional constraints on the planetary atmospheres. Another way to improve the parameter retrieval is to obtain mid-infrared spectra with even higher spectral resolution and S/N. In a recent qualitative study, \cite{Leger:2015} suggest that a spectral resolution of 40 and a S/N of 20 are actually required to unambiguously detect CO$_2$, O$_3$ and H$_2$O in the spectrum of an Earth-like planet. While this needs to be confirmed by a detailed analysis similar to that performed by von Paris et al.\cite{vonParis:2013}, one can see in Figure~\ref{fig:case_out} that these requirements are also sufficient to resolve all spectral features (i.e., H$_2$O, O$_3$, CO$_2$, CH$_4$) of the test cases described above and presented in Figure~\ref{fig:case_out}, except for O$_3$ in the atmosphere of the water ocean planet, which would require a higher S/N or follow-up observations. Achieving such a high S/N is crucial to break model degeneracies inherent to parameter retrieval and will drive the integration time spent on each target for a given collecting area (and hence the number of planetary atmospheres than can be studied in a given mission lifetime).


\section{SUMMARY AND CONCLUSIONS} \label{sec:conclusions}  

A space-based nulling interferometer with four 75-cm collecting apertures and a minimum baseline of 40 m could take a mid-infrared (5-20 microns) medium-resolution (R=40) spectrum of Proxima b in one day of integration. Such a spectrum would reveal whether the planet has an atmosphere and would constrain its surface temperature and pressure as well as reveal the presence of important atmospheric species such as CO$_2$, H$_2$O, and, in some cases, O$_3$. Most technologies necessary to build such an instrument have benefited from intense research in the past and have today reached TRL 5. A relatively small instrument dedicated to Proxima b will both lead to transformational science and serve as crucial stepping stone for a future flagship instrument that will be able to characterize a large sample of rocky exoplanets (see Figure~\ref{fig:synergies} and more quantitative predictions in Kammerer and Quanz 2018\cite{Kammerer:2018}).

\acknowledgments 

DD and OA thank the Belgian national funds for scientific research (FNRS). This work was partly funded by the European Research Council under the European Union's Seventh Framework Program (ERC Grant Agreement n. 337569) and by the French Community of Belgium through an ARC grant for Concerted Research Action. 


\bibliographystyle{spiebib} 

\begin{thebibliography}{10}

\bibitem{Mayor:1995}
{Mayor}, M. and {Queloz}, D., ``{A Jupiter-mass companion to a solar-type
  star},'' {\em \nat}~{\bf 378},  355--359 (Nov. 1995).

\bibitem{Winn:2015}
{Winn}, J.~N. and {Fabrycky}, D.~C., ``{The Occurrence and Architecture of
  Exoplanetary Systems},'' {\em \araa}~{\bf 53},  409--447 (Aug. 2015).

\bibitem{Anglada:2016}
Anglada-Escud{\'e}, G., Amado, P.~J., Barnes, J., Berdi{\~n}as, Z.~M., Butler,
  R.~P., Coleman, G. A.~L., de~la Cueva, I., Dreizler, S., Endl, M., Giesers,
  B., Jeffers, S.~V., Jenkins, J.~S., Jones, H. R.~A., Kiraga, M., K{\"u}rster,
  M., L{\'o}pez-Gonz{\'a}lez, M.~J., Marvin, C.~J., Morales, N., Morin, J.,
  Nelson, R.~P., Ortiz, J., Ofir, A., Paardekooper, S.-J., Reiners, A.,
  Rodr{\'\i}guez, E., Rodrίguez-L{\'o}pez, C., Sarmiento, L.~F., Strachan,
  J.~P., Tsapras, Y., Tuomi, M., and Zechmeister, M., ``A terrestrial planet
  candidate in a temperate orbit around proxima centauri,'' {\em Nature}~{\bf
  536},  437--440 (08 2016).

\bibitem{Turbet:2016}
{Turbet}, M., {Leconte}, J., {Selsis}, F., {Bolmont}, E., {Forget}, F.,
  {Ribas}, I., {Raymond}, S.~N., and {Anglada-Escud{\'e}}, G., ``{The
  habitability of Proxima Centauri b II. Possible climates and
  Observability},'' {\em ArXiv e-prints}  (Aug. 2016).

\bibitem{Turbet:2018}
{Turbet}, M., {Bolmont}, E., {Leconte}, J., {Forget}, F., {Selsis}, F.,
  {Tobie}, G., {Caldas}, A., {Naar}, J., and {Gillon}, M., ``{Modeling climate
  diversity, tidal dynamics and the fate of volatiles on TRAPPIST-1 planets},''
  {\em \aap}~{\bf 612},  A86 (May 2018).

\bibitem{Meadows:2018}
{Meadows}, V.~S., {Arney}, G.~N., {Schwieterman}, E.~W., {Lustig-Yaeger}, J.,
  {Lincowski}, A.~P., {Robinson}, T., {Domagal-Goldman}, S.~D., {Deitrick}, R.,
  {Barnes}, R.~K., {Fleming}, D.~P., {Luger}, R., {Driscoll}, P.~E., {Quinn},
  T.~R., and {Crisp}, D., ``{The Habitability of Proxima Centauri b:
  Environmental States and Observational Discriminants},'' {\em
  Astrobiology}~{\bf 18},  133--189 (Feb. 2018).

\bibitem{Kreidberg:2016}
{Kreidberg}, L. and {Loeb}, A., ``{Prospects for Characterizing the Atmosphere
  of Proxima Centauri b},'' {\em \apjl}~{\bf 832},  L12 (Nov. 2016).

\bibitem{GomezLeal2016}
{G{\'o}mez-Leal}, I., {Codron}, F., and {Selsis}, F., ``{Thermal light curves
  of Earth-like planets: 1. Varying surface and rotation on planets in a
  terrestrial orbit},'' {\em \icarus}~{\bf 269},  98--110 (May 2016).

\bibitem{ThesisIlleana}
{G{\'o}mez-Leal}, I., {\em Spectrophotometry of the infrared emission of
  Earth-like Planets (PhD Thesis)}, PhD thesis, University of Bordeaux, France
  (2013).

\bibitem{Desmarais:2002}
{Des Marais}, D.~J., {Harwit}, M.~O., {Jucks}, K.~W., {Kasting}, J.~F., {Lin},
  D.~N.~C., {Lunine}, J.~I., {Schneider}, J., {Seager}, S., {Traub}, W.~A., and
  {Woolf}, N.~J., ``{Remote Sensing of Planetary Properties and Biosignatures
  on Extrasolar Terrestrial Planets},'' {\em Astrobiology}~{\bf 2},  153--181
  (June 2002).

\bibitem{vonParis:2013}
{von Paris}, P., {Hedelt}, P., {Selsis}, F., {Schreier}, F., and {Trautmann},
  T., ``{Characterization of potentially habitable planets: Retrieval of
  atmospheric and planetary properties from emission spectra},'' {\em
  \aap}~{\bf 551},  A120 (Mar. 2013).

\bibitem{Lord:1992}
{Lord}, S.~D., ``{A new software tool for computing Earth's atmospheric
  transmission of near- and far-infrared radiation},'' tech. rep. (Dec. 1992).

\bibitem{Colavita:2009}
{Colavita}, M.~M., {Serabyn}, E., {Millan-Gabet}, R., {Koresko}, C.~D.,
  {Akeson}, R.~L., {Booth}, A.~J., {Mennesson}, B.~P., {Ragland}, S.~D.,
  {Appleby}, E.~C., {Berkey}, B.~C., {Cooper}, A., {Crawford}, S.~L.,
  {Creech-Eakman}, M.~J., {Dahl}, W., {Felizardo}, C., {Garcia-Gathright},
  J.~I., {Gathright}, J.~T., {Herstein}, J.~S., {Hovland}, E.~E., {Hrynevych},
  M.~A., {Ligon}, E.~R., {Medeiros}, D.~W., {Moore}, J.~D., {Morrison}, D.,
  {Paine}, C.~G., {Palmer}, D.~L., {Panteleeva}, T., {Smith}, B., {Swain},
  M.~R., {Smythe}, R.~F., {Summers}, K.~R., {Tsubota}, K., {Tyau}, C.,
  {Vasisht}, G., {Wetherell}, E., {Wizinowich}, P.~L., and {Woillez}, J.~M.,
  ``{Keck Interferometer Nuller Data Reduction and On-Sky Performance},'' {\em
  \pasp}~{\bf 121},  1120--1138 (Oct. 2009).

\bibitem{Defrere:2016}
{Defr{\`e}re}, D., {Hinz}, P.~M., {Mennesson}, B., {Hoffmann}, W.~F.,
  {Millan-Gabet}, R., {Skemer}, A.~J., {Bailey}, V., {Danchi}, W.~C., {Downey},
  E.~C., {Durney}, O., {Grenz}, P., {Hill}, J.~M., {McMahon}, T.~J., {Montoya},
  M., {Spalding}, E., {Vaz}, A., {Absil}, O., {Arbo}, P., {Bailey}, H.,
  {Brusa}, G., {Bryden}, G., {Esposito}, S., {Gaspar}, A., {Haniff}, C.~A.,
  {Kennedy}, G.~M., {Leisenring}, J.~M., {Marion}, L., {Nowak}, M., {Pinna},
  E., {Powell}, K., {Puglisi}, A., {Rieke}, G., {Roberge}, A., {Serabyn}, E.,
  {Sosa}, R., {Stapeldfeldt}, K., {Su}, K., {Weinberger}, A.~J., and {Wyatt},
  M.~C., ``{Nulling Data Reduction and On-sky Performance of the Large
  Binocular Telescope Interferometer},'' {\em \apj}~{\bf 824},  66 (June 2016).

\bibitem{Kendrew:2008}
{Kendrew}, S., {Jolissaint}, L., {Mathar}, R.~J., {Stuik}, R., {Hippler}, S.,
  and {Brandl}, B., ``{Atmospheric refractivity effects on mid-infrared ELT
  adaptive optics},'' in [{\em Adaptive Optics
  Systems}{\nolinebreak\hspace{0.1em}]},  {\em Proc. SPIE} {\bf 7015},  70155T
  (July 2008).

\bibitem{Kaltenegger:2010}
{Kaltenegger}, L., {Eiroa}, C., and {Fridlund}, C.~V.~M., ``{Target star
  catalogue for Darwin Nearby Stellar sample for a search for terrestrial
  planets},'' {\em \apss}~{\bf 326},  233--247 (Apr. 2010).

\bibitem{Brandl:2014}
{Brandl}, B.~R., {Feldt}, M., {Glasse}, A., {Guedel}, M., {Heikamp}, S.,
  {Kenworthy}, M., {Lenzen}, R., {Meyer}, M.~R., {Molster}, F., {Paalvast}, S.,
  {Pantin}, E.~J., {Quanz}, S.~P., {Schmalzl}, E., {Stuik}, R., {Venema}, L.,
  and {Waelkens}, C., ``{METIS: the mid-infrared E-ELT imager and
  spectrograph},'' in [{\em Ground-based and Airborne Instrumentation for
  Astronomy V}{\nolinebreak\hspace{0.1em}]},  {\em Proc.\ SPIE} {\bf 9147},
  914721 (Aug. 2014).

\bibitem{Burrows:2014}
{Burrows}, A.~S., ``{Highlights in the study of exoplanet atmospheres},'' {\em
  \nat}~{\bf 513},  345--352 (Sept. 2014).

\bibitem{Boccaletti:2015}
{Boccaletti}, A., {Lagage}, P.-O., {Baudoz}, P., {Beichman}, C., {Bouchet}, P.,
  {Cavarroc}, C., {Dubreuil}, D., {Glasse}, A., {Glauser}, A.~M., {Hines},
  D.~C., {Lajoie}, C.-P., {Lebreton}, J., {Perrin}, M.~D., {Pueyo}, L.,
  {Reess}, J.~M., {Rieke}, G.~H., {Ronayette}, S., {Rouan}, D., {Soummer}, R.,
  and {Wright}, G.~S., ``{The Mid-Infrared Instrument for the James Webb Space
  Telescope, V: Predicted Performance of the MIRI Coronagraphs},'' {\em
  \pasp}~{\bf 127},  633--645 (July 2015).

\bibitem{Brandl:2016}
Brandl, B.~R., Agocs, T., Aitink-Kroes, G., Bertram, T., Bettonvil, F., van
  Boekel, R., Boulade, O., Feldt, M., Glasse, A., Glauser, A., Godel, M.,
  Hurtado, N., Jager, R., Kenworthy, M.~A., Mach, M., Meisner, J., Meyer, M.,
  Pantin, E., Quanz, S., Schmid, H.~M., Stuik, R., Veninga, A., and Waelkens,
  C. (2016).

\bibitem{Quanz15}
{Quanz}, S.~P., {Crossfield}, I., {Meyer}, M.~R., {Schmalzl}, E., and {Held},
  J., ``{Direct detection of exoplanets in the 3-10 {$\mu$}m range with
  E-ELT/METIS},'' {\em International Journal of Astrobiology}~{\bf 14},
  279--289 (Apr. 2015).

\bibitem{Bracewell:1978}
{Bracewell}, R.~N., ``{Detecting nonsolar planets by spinning infrared
  interferometer},'' {\em \nat}~{\bf 274},  780 (Aug. 1978).

\bibitem{Lay:2004}
{Lay}, O.~P., ``{Systematic Errors in Nulling Interferometers},'' {\em
  \ao}~{\bf 43},  6100--6123 (Nov. 2004).

\bibitem{Angel:1997}
{Angel}, J.~R., {Burge}, J.~H., and {Woolf}, N.~J., ``{Detection and
  spectroscopy of exo-planets like Earth},'' in [{\em Optical Telescopes of
  Today and Tomorrow}{\nolinebreak\hspace{0.1em}]},  {Ardeberg}, A.~L., ed.,
  {\em Proceedings SPIE} {\bf 2871},  516--519 (Mar. 1997).

\bibitem{Mennesson:1997}
{Mennesson}, B. and {Mariotti}, J.~M., ``{Array Configurations for a Space
  Infrared Nulling Interferometer Dedicated to the Search for Earthlike
  Extrasolar Planets},'' {\em \icarus}~{\bf 128},  202--212 (July 1997).

\bibitem{Defrere:2010}
{Defr{\`e}re}, D., {Absil}, O., {den Hartog}, R., {Hanot}, C., and {Stark}, C.,
  ``{Nulling interferometry: impact of exozodiacal clouds on the performance of
  future life-finding space missions},'' {\em \aap}~{\bf 509},  A9 (Jan. 2010).

\bibitem{Leger:1996}
{L{\'e}ger}, A., {Mariotti}, J.~M., {Mennesson}, B., {Ollivier}, M., {Puget},
  J.~L., {Rouan}, D., and {Schneider}, J., ``{The DARWIN project},'' {\em
  \apss}~{\bf 241},  135--146 (Mar. 1996).

\bibitem{Lay:2005}
{Lay}, O.~P., ``{Imaging properties of rotating nulling interferometers},''
  {\em \ao}~{\bf 44},  5859--5871 (Oct. 2005).

\bibitem{Defrere:2017}
{Defr{\`e}re}, D., {Absil}, O., and {Beichman}, C.,  [{\em {Interferometric
  Space Missions for Exoplanet Science: Legacy of
  Darwin/TPF}}{\nolinebreak\hspace{0.1em}]},  82 (2017).

\bibitem{Peters:2010}
{Peters}, R.~D., {Lay}, O.~P., and {Lawson}, P.~R., ``{Mid-Infrared Adaptive
  Nulling for the Detection of Earthlike Exoplanets},'' {\em \pasp}~{\bf 122},
  85--92 (Jan. 2010).

\bibitem{Martin:2010}
{Martin}, S.~R. and {Booth}, A.~J., ``{Demonstration of exoplanet detection
  using an infrared telescope array},'' {\em \aap}~{\bf 520},  A96 (Sept.
  2010).

\bibitem{Mennesson:2005}
{Mennesson}, B., {L{\'e}ger}, A., and {Ollivier}, M., ``{Direct detection and
  characterization of extrasolar planets: The Mariotti space interferometer},''
  {\em Icarus}~{\bf 178},  570--588 (Nov. 2005).

\bibitem{Rauer:2011}
{Rauer}, H., {Gebauer}, S., {Paris}, P.~V., {Cabrera}, J., {Godolt}, M.,
  {Grenfell}, J.~L., {Belu}, A., {Selsis}, F., {Hedelt}, P., and {Schreier},
  F., ``{Potential biosignatures in super-Earth atmospheres. I. Spectral
  appearance of super-Earths around M dwarfs},'' {\em \aap}~{\bf 529},  A8 (May
  2011).

\bibitem{Tian:2014}
{Tian}, F., {France}, K., {Linsky}, J.~L., {Mauas}, P.~J.~D., and {Vieytes},
  M.~C., ``{High stellar FUV/NUV ratio and oxygen contents in the atmospheres
  of potentially habitable planets},'' {\em Earth and Planetary Science
  Letters}~{\bf 385},  22--27 (Jan. 2014).

\bibitem{Leger:2004}
{L{\'e}ger}, A., {Selsis}, F., {Sotin}, C., {Guillot}, T., {Despois}, D.,
  {Mawet}, D., {Ollivier}, M., {Lab{\`e}que}, A., {Valette}, C., {Brachet}, F.,
  {Chazelas}, B., and {Lammer}, H., ``{A new family of planets?
  ``Ocean-Planets''},'' {\em \icarus}~{\bf 169},  499--504 (June 2004).

\bibitem{Selsis:2002}
{Selsis}, F., {Despois}, D., and {Parisot}, J.-P., ``{Signature of life on
  exoplanets: Can Darwin produce false positive detections?},'' {\em \aap}~{\bf
  388},  985--1003 (June 2002).

\bibitem{Leger:2015}
{L{\'e}ger}, A., {Defr{\`e}re}, D., {Malbet}, F., {Labadie}, L., and {Absil},
  O., ``{Impact of {$\eta$}$_{Earth}$ on the Capabilities of Affordable Space
  Missions to Detect Biosignatures on Extrasolar Planets},'' {\em \apj}~{\bf
  808},  194 (Aug. 2015).

\bibitem{Kammerer:2018}
{Kammerer}, J. and {Quanz}, S.~P., ``{Simulating the exoplanet yield of a
  space-based mid-infrared interferometer based on Kepler statistics},'' {\em
  \aap}~{\bf 609},  A4 (Jan. 2018).

\end{thebibliography}

\end{document}